# Microstructural and morphological properties of homoepitaxial (001)ZnTe layers investigated by x-ray diffuse scattering

T. Di Luccio, G. Scalia, and L. Tapfer

*Ente per le Nuove tecnologie, l'Energia e l'Ambiente (ENEA), UTS-MAT, Centro Ricerche Brindisi, Strada Statale 7 Appia, km. 713+700, I-72100 Brindisi, Italy*

P. Morales

*Ente per le Nuove tecnologie, l'Energia e l'Ambiente (ENEA), UTS-MAT, Centro Ricerche Casaccia, Via Anguillarese 301, I-00060 S. Maria di Galeria, Italy*

M. Traversa

*Dipartimento di Ingegneria dell'Innovazione, Università di Lecce, Via Arnesano, I-73100 Lecce, Italy*

P. Prete

*Istituto per la Microelettronica e i Microsistemi (IMM) del CNR, Sez. di Lecce, Via Arnesano, I-73100 Lecce, Italy*

N. Lovergine [a]

*Dipartimento di Ingegneria dell'Innovazione, Università di Lecce, Via Arnesano, I-73100 Lecce, Italy*

**ABSTRACT**

The microstructural and morphological properties of homoepitaxial (001)ZnTe layers grown by metalorganic vapor phase epitaxy at a temperature ($T_G$) between 325 °C and 400 °C are investigated by x-ray diffuse scattering. High resolution reciprocal space maps (RSMs) recorded close to the ZnTe (004) Bragg peak show different diffuse scattering features that can be ascribed to (i) the specific surface morphology of the sample, and (ii) the presence of extended lattice defects in the epilayers. One kind of cross-shaped diffuse scattering streaks, appearing for $T_G \geq 350$ °C, extended along <111> directions and can be attributed to stacking faults (SFs) occurring at the epilayer-substrate interface, within the epilayers. The SF diameter was estimated around 200-300 nm, while their density increases with $T_G$. Another kind of cross-shaped diffuse streaks, inclined at an angle $\beta \approx 80°$ with respect to the <110> in-plane direction, arises from the morphology of epilayers grown above 360°C, their surfaces being covered by pairs of pyramidal hillocks up to a density of $10^6 - 10^7$ cm$^{-2}$. Atomic force microscopy (AFM) measurements showed that the apex angles of the pyramids compare well with the value of $2\beta$. The hillock formation is ascribed to Te adatoms experiencing a Schwoebel potential barrier at the step edges around pairs of partial dislocations (dipoles) bounding the SFs. In a quite narrow growth temperature interval around 350°C no $\beta$-crossed diffuse streaks are instead observed in the RSMs, indicating a smooth ZnTe surface. Finally, at a lower growth temperature ($T_G$=325°C) a diffuse scattering intensity distribution defined by an angle $\gamma \approx 63°$ with respect to the <110> in-plane direction is observed, corresponding to a dense ridging of the epilayer surface along the perpendicular direction. Both RSM analysis and AFM measurements indicate that the ridge sidewalls are {113} planes.

PACS nos.:

68.55.-a    61.72.Dd    68.37.-d    81.15.Gh

[a] corresponding author: nico.lovergine@unile.it

## I. INTRODUCTION

Light emitting diodes (LED) and laser diodes (LD) operating in the red, green, blue and ultraviolet (UV) spectral regions have up to now employed mostly III-V and III-N compound semiconductors. In the past much attention has been devoted to the application of wide band-gap II-VI compound semiconductors, such as ZnSe and ZnCdSe alloys, as advantageous materials for the realization of high efficiency LEDs and LDs operating in the blue-green region.[1] Since recently ZnO and related alloys are attracting great interest for opto-electronic devices emitting in the blue-UV wavelength region.[2] Among other II-VI semiconductors, ZnTe [with a direct band-gap of 2.26 eV (548.5 nm) at 300 K] and Zn-rich telluride alloys, such as ZnCdTe, are instead particularly promising for the realization of high brightness LEDs and LDs operating in the green and green-yellow spectral range (520-580 nm) of light.

Unfortunately, the lack of high quality II-VI single crystal substrates and the difficult *p* type doping of most wide gap II-VI compounds have prevented until now the fabrication of long lifetime LEDs and LDs. In the past, the epitaxial growth of II-VI based devices was indeed attempted on GaAs substrates. However, the chemical, lattice and thermal mismatches between II-VI epilayers and GaAs lead to residual lattice strains and interface defects, among the major factors limiting device performances.[3,4] Similarly, the epitaxy of ZnO and its alloys is performed on highly mismatched sapphire substrates[5] or on hybrid GaN/Si structures.[6]

In the case of ZnTe-based devices the substrate limitations have been recently overcome as high crystalline quality ZnTe wafers obtained by the vertical gradient freezing (VGF) method[7,8] became available, allowing the homoepitaxy of good structural and optical quality ZnTe by both molecular beam (MBE)[9] and metalorganic vapor phase (MOVPE) epitaxy.[10] Moreover, while VGF-grown ZnTe substrates can be easily made *p* type by phosphorous (P) doping, achieving net hole concentrations up to $10^{18}$ cm$^{-3}$, *n* type ZnTe:Al homoepitaxial layers with electron concentrations up to $4\times10^{18}$ cm$^{-3}$ were demonstrated by MBE.[11] This opened the way to the realization of ZnTe-based vertical p-i-n diode structures emitting between 520 nm and 578 nm.[12,13] An advantage of such vertical structure over GaN-based devices (deposited on highly insulating sapphire) is that it avoids further processing steps for the lateral electrical contacting. The growth of ZnTe-based devices on ZnTe substrates has thus technological potentials for applications to TV projectors and signal transmission (at around 560 nm) through polymer optical fibers, as alternatives to both low quantum efficiency GaP- (Ref. 14) and complex InGaN-based green LEDs and LDs.

Within this scenario it is of primary importance to further optimize the growth of ZnTe homoepitaxial layers by studying their microstructural and morphological properties.

The use of high-angle X-ray diffuse scattering (XDS) experiments has been often reported in the literature to reveal different kinds of lattice defects in heteroepitaxial structures, such as point defects,[15,16] dislocations,[17-19] misfit dislocation at heterointerfaces,[20-22] stacking faults[23-25] and extended lattice defects.[23,25] In addition, XDS was also employed with success to study artificial periodic surface gratings[18] or laterally modulated structures.[19,20]

In this paper we report on the XDS study of homoepitaxial ZnTe layers deposited by MOVPE on VGF-grown (100)ZnTe substrates. We present high resolution reciprocal space map (HR-RSM) measurements performed on relatively thick ZnTe epilayers grown at different temperatures. The study of X-ray diffuse scattering in HR-RSMs allows us to investigate both lattice and morphological imperfections of the material. The epilayer morphological properties drawn from the analysis of HR-RSMs are compared with and further supported by direct observations of the sample surface by atomic force microscopy (AFM) and field emission gun scanning electron microscopy (FEG-SEM). It is also demonstrated that reciprocal space mapping allows an efficient and highly sensitive assessment of the epilayer microstructural properties. The results are discussed in terms of fundamental growth mechanisms.

## II. EXPERIMENT

Homoepitaxial ZnTe layers were deposited on (001)ZnTe:P substrates by atmospheric pressure MOVPE using dimethylzinc (Me$_2$Zn) and di-isopropyltelluride ($^i$Pr$_2$Te) as Zn and Te precursors, respectively.[10,21] The substrates were cleaved from ZnTe:P wafers supplied by Nikko Materials Co., Ltd. of Japan. In order to eliminate any oxide or contaminants from their surface and favor the epitaxy of ZnTe, as-received substrates were carefully cleaned and etched as described previously.[10,21] The substrates were then heat-treated in the MOVPE reactor under 1.0 l/min H$_2$ flow at the optimal[21] temperature of 350°C immediately before the epilayer growth. For the present work, several 3 μm thick epilayer samples were deposited at temperatures ($T_G$) of 325°C, 350°C, 375°C and 400°C, the Te:Zn precursors molar flow ratio in the vapor being kept fixed at 1:1.

For the X-ray diffraction experiments a Philips X'Pert Pro instrument was employed. As monochromator-collimator system a parabolic multilayer mirror and a four-bounce Bartels monochromator [Ge-crystals and (220) reflections] was used. The sample was mounted on an Eulerian cradle sample stage with independent movements for the incident angle $\omega$, the diffraction angle $2\theta$, the azimuth angle $\phi$ and the tilt angle $\psi$. In triple axis mode a channel-cut Ge crystal analyzer in front of the detector reduces the angular acceptance to 12 arcsec or less, making the instrument suitable for high resolution mapping of the reciprocal lattice. RSMs in the vicinity of the ZnTe (004) reciprocal lattice point were recorded with area scans over 1° range along the $\omega/2\theta$ – axis and 0.5° range along $\omega$ – axis in about 18 hours. Rocking curves were also performed in order to investigate the crystalline quality of the epilayer with respect to the substrate.

Noteworthy to comment here is that we revealed the particular XDS features reported in the high angle reciprocal space mapping of this work only by using the Philips X'Pert Pro instrument, while the same measurements performed on a Philips MRD instrument did not provide the same clear information due to its reduced dynamic intensity range. In fact the background intensity of the X'Pert Pro is about one order of magnitude lower with respect to the MRD due to the curved multilayer mirror used for the focusing of the primary X-ray beam.

The epilayer surface morphology was observed by using a FEG-SEM microscope (JEOL model JSM 6500 F) operating with a primary electron beam energy of 5kV. Contact mode AFM was also used to quantitatively investigate the surface of the samples. In order to obtain fine details of the epilayer morphology, a Quesant Nomad microscope was equipped with very thin and sharp silicon pyramidal tips, their radius of curvature being about 20 nm, as evaluated by SEM.

## III. MICROSTRUCTURAL AND MORPHOLOGICAL ANALYSES

### A. X-ray reciprocal space mapping

High-resolution X-ray rocking curves were recorded for different azimuth angles in symmetrical diffraction geometry around the (004) and (006) reciprocal lattice points (relp), and in asymmetrical diffraction arrangement around the (115) and (224) relp. These measurements showed that all epilayers are stress-free and, in addition, no tilt of the epilayer with respect to the substrate could be revealed. Moreover, the long range crystallinity of the epilayers compares well with that of the substrates.[21]

Figures 1 and 2 show the sample HR-RSMs recorded in the vicinity of the (004) relp and (Fig. 2) for two different azimuth angles ($\phi=0°$ and $\phi=90°$). The azimuth angle positions $\phi=0°$ and $\phi=90°$ correspond to the sample setting (sample rotation around the surface normal by an angle $\phi$) with the $[110]$- and $[\bar{1}10]$-directions lying in the scattering plane. All maps in the figures are reported in reciprocal lattice units, with the horizontal axis representing the in-plane scattering vector component $Q_x$ along one of the <110> directions and the vertical axis representing the scattering vector component $Q_y$ perpendicular to the epilayer surface ([001] direction). The horizontal and vertical scales and the intensity contour levels are kept equal for all maps.

Figure 1(a) shows the HR-RSM recorded from a ZnTe homoepitaxial layer grown at $T_G$ =350°C. A diffuse scattering that is more pronounced along certain directions (indicated by the angle α) in the reciprocal space is well observed, giving rise to a cross-shaped intensity contour map. For comparison the HR-RSM of the as-etched ZnTe:P substrate is reported in Fig.1(b), showing an almost circularly-shaped diffuse scattering halo around the (004) relp, i.e. without any noticeable diffuse scattering feature.

In Fig.2 we report the HR-RSMs measured for the samples grown at 325°C, 375°C and 400°C at the two azimuth angles ϕ=0° and ϕ=90°. All HR-RSMs are characterized by diffuse scattering halos that are more pronounced along certain directions of the reciprocal space. In particular, for the epilayers grown at 375°C and 400°C [Fig. 2(b) and 2(c), respectively] two distinct cross-shaped features (streaks) arise, one having an aperture angle $2\alpha \approx 70°$ and the other an aperture angle $2\beta \approx 160°$, where α and β are the angles between the diffuse scattering streak directions and the <110> direction ($Q_x$-axis). Notably, the aperture angle α reported in Fig. 1(a) for the 350°C sample coincides, within experimental errors, with the one in Fig.2(b) and 2(c). Moreover, while the α–crossed streaks appear in all the samples but the one grown at $T_G$ = 325°C [Fig. 2(a)], the β–crossed streaks are observed only above 360°C. Also, the β–crossed streaks are observed at both ϕ=0° and ϕ=90° for the 400°C sample [Fig. 2(c)], whereas the epilayer grown at $T_G$ =375°C does not show those streaks for ϕ=90° [Fig. 2(b)].

Finally, the HR-RSMs recorded at ϕ=0° and ϕ=90° for the epilayer grown at $T_G$ =325 °C show a clear difference with respect to the other maps. In fact, the maps in Fig. 2(a) do not show either of the cross-shaped streaks above; instead, a peculiar diffuse scattering halo can be observed for one specific sample setting, i.e. ϕ=90°. This halo seems more pronounced in directions defined by an aperture angle $2\gamma \approx 126°$ [see Fig. 2(a)], where the angle γ is again defined with respect to the sample <110> in-plane-direction.

It is well known that the XDS intensity distribution in high-angle RSMs can be influenced by different factors related to the materials structure and morphology. In particular, the shape of the diffusely scattered intensity distribution around a relp can be affected by structural defects (dislocations, point defects, clusters, etc.) or by morphological peculiarities (surface roughness, islands, 3D growth, etc.). Therefore, in order to discriminate between morphological and structural induced features in our HR-RSMs and to make a correct attribution of experimental observations we analyzed the present ZnTe epilayers also by high-resolution surface imaging techniques, like FEG-SEM and AFM.

**B. Epilayer surface morphology characterization**

The morphological properties of ZnTe epilayers show a strong dependence on the growth temperature.[21] A nearly featureless surface morphology is observed only for epilayers grown within a limited temperature interval around 350 °C, corresponding to the transition between surface kinetics and mass transport limited growth, while samples grown below or above this interval show a distinctive surface morphology. Figure 3 shows FEG-SEM images of the surfaces of three samples grown at 325°C, 350°C, and 375°C. While the sample grown at 350°C shows a very smooth surface [Fig.3(b)], a well pronounced asymmetry (ridging) of the surface texture along one of the equivalent <110> directions is observed for the sample grown at 325°C [Fig.3(a)]. We previously demonstrated that such ridging is ascribable to the development of {113}Te facets, the slowest to grow under surface kinetics limited conditions.[22] Large pyramidal hillocks appear on the surface of epilayers grown at $T_G$ >360°C.[21] Figure 3(c) shows that for $T_G$ =375°C the hillocks are elongated rhombic pyramidal structures with the major axis of the rhombus-base parallel to the <110> direction. The hillocks are packed more densely in one of <110>-directions, forming chains aligned along the preferential direction. As the temperature is raised to 400°C the pyramids become very distinct, larger and cover uniformly the surface.[21]

AFM measurements performed on the 350°C sample (not reported here) confirmed the good morphology of the epilayer surface, whose root mean square roughness ranges around 1.05 nm, i.e. lower than that of the substrate (5.6 nm).

Figure 4(a) shows an AFM topographic image taken over a 10×10 µm$^2$ surface area of the 325°C epilayer; the sample dense surface ridging along a <110> direction clearly appears with an average ridge distance of about 1µm, while in the perpendicular direction the ridges extend to several µm. A line scan normal to the ridges reveals the surface profile [Fig. 4(b)] and allows us to estimate the inclination angle of the ridge sidewalls, whose average value turns out to be 66°±4°. This inclination angle compares well, within experimental errors, with the 64.76° theoretical value of the angle between [113]- and [001]-directions in ZnTe, further confirming what reported in Ref.22. Notably, Fig.4(b) also shows that the peak-to-valley height variations of the epilayer ridged surface exceeds 200 nm.

Fig. 5(a) shows an AFM topographic image taken over a 3×3 µm$^2$ surface area of the sample grown at 375°C. The image clearly shows the presence of pyramidal hillocks as in Fig.3(c); to be able to analyze these structures we approximated them as perfect pyramids even if some of them exhibit multiple faces or incomplete growth. We thus identified two directions ($\phi=0°$ and $\phi=90°$ in the figure) corresponding to the major and minor diagonals of the hillock rhombus-base, respectively. According to the line scan height profiles in Fig.5(b), recorded along $\phi=0°$ and $\phi=90°$, the length of the rhombus-base axes of the hillocks ranges from several hundreds nm up to a micron, while the height of the pyramids is about 100 nm. However, the FEG-SEM micrograph of Fig.3(c) indicates that even bigger hillocks can occur randomly on the epilayer surface. Still several AFM line scans performed at different location across the sample surface allowed us to determine the average values of the apex angles of the pyramids, which turned out to be 172°±2° along $\phi=0°$ and 162°±2° along $\phi=90°$. Similar results were obtained for the sample grown at 400°C; despite the hillocks are more extended in size,[21] the dimensions of the rhombus-base of the pyramids are in the micron range, their heights reaching over 1μm. In this case the average apex angles of the pyramids obtained from AFM line scans in the ϕ=0° and ϕ=90° directions equal 165°±2° and 166°±2°, respectively.

## IV. INTERPRETATION OF RECIPROCAL SPACE MAPPING: EPILAYER SURFACE MORPHOLOGY AND LATTICE DEFECTS

As evidenced by the experimental X-ray diffraction measurements in Sec. III.A, different features appear in the HR-RSMs of the epilayers, the XDS intensity being more pronounced along specific crystallographic directions of the reciprocal space. In the following we clarify the origin of such XDS features on the basis of morphological analyses (Sec. III.B) and previous studies in the literature.

We recall that for the $\alpha$-crossed streaks [Fig.1(a) and Figs.2(b) and 2(c)] $\alpha \approx 35°$, thus coinciding well with the theoretical value (35.26°) of the angle between {111} and {001} planes of the cubic lattice. Therefore, a <111> direction can be assigned to one of the $\alpha$-crossed streak ($<\bar{1}\bar{1}1>$ for the other streak). Notably, these streaks were detected independently of the epilayer surface morphology: in fact, the epilayer grown at 350°C shows weak $\alpha$-crossed streaks despite its surface morphology [Fig.3(b)] is very different from that of the other samples. Moreover, the streaks are not observed for the ZnTe:P substrate, which surface is comparable to that of the 350°C sample.[21] Instead the substrate HR-RSM is characterized by a circularly-shaped diffuse scattering halo, as expected for an almost perfect (low defect density) crystal. This clearly indicates that the $\alpha$-crossed streaks are not related to specific surface morphology features, but should be instead ascribed to intrinsic lattice defects present in the ZnTe epilayers.

Theoretical calculations of the X-ray diffracted intensity from several defects randomly distributed in crystals have been carried out by Holy and Kubena.[19] In the case of stacking faults (SFs) on {111} planes the RSMs have the same shape we observe in Fig.1(a) and Figs.2(b) and 2(c). Similar diffuse scattering streaks were also observed along <111> directions in boron implanted Si samples by grazing incidence RSM measurements.[15] The origin of the streaks was attributed to extrinsic SFs by Larson et al.,[24] the SF diameter perpendicular to the {111} planes being evaluated from the width of the diffuse scattering streaks.

Despite the different measurement geometry, the information we get from the HR-RSMs is comparable to what found in Ref. 15. Therefore, we ascribe the $\alpha$-crossed streaks observed in Fig.1(a) and Fig.2(b) and 2(c) to SFs lying along {111} planes within the ZnTe epilayers. Figures 6(a) and 6(b) report a scheme of the SF scattering geometry in both the direct and reciprocal space. An estimate of the streak width in Fig. 2 gives diameters of the SFs perpendicular to the {111} planes comprised between 200 nm and 300 nm.

$\beta$-crossed streaks appear in the HR-RSMs of epilayers grown above 375°C (Sec. III.A), while they do not appear for the samples grown at 350°C or below, i.e. irrespective whether SFs occur or not into these layers. The streaks seems thus to arise from the morphological features of the epilayers grown above 350°C. Indeed, $\beta$-crossed streaks become more and more pronounced at higher values of $T_G$ when pyramidal hillocks begin to cover the epilayer surface. The angle $\beta$ observed in the HR-RSMs should thus correspond to the average inclination angle of the hillock edges/facets with respect to the (001) plane (parallel to the substrate surface). A scheme of the scattering geometry from such a hillock-covered surface (defined by a pyramid apex angle 2β) is shown in Figs.6(c) and 6(d) in the direct and reciprocal space, respectively.

In order to assess quantitatively the above diffraction scheme, we compared the aperture angle $2\beta$ measured in the HR-RSMs of Fig. 2 and the value of the average pyramid apex angle measured by AFM line scans (Sec. III.B). As a matter of fact the $\beta$-crossed streaks in the HR-RSMs of the 400°C epilayer [Fig.2(c)] show a total aperture angle $2\beta \approx 170°$ along $\phi=0°$ and $2\beta \approx 162°$ along $\phi= 90°$, which compare quite well with the 165°±2° and 166°±2° values, respectively determined by the AFM line scans (Sec. III.B). Since the information of RSMs is averaged over an area that is about $10^6$ larger with respect the information obtained from AFM images, we can assume that the XDS yield more accurate average data. Somehow different is the case of the epilayer grown at $T_G$ =375°C. Here the morphology is strongly directional along $\phi$=0° [Figs. 3(c) and 5(a)] where the pyramids merge into one another, while in the perpendicular direction the density is clearly lower. This is also confirmed by the asymmetry observed in the sample HR-RSMs [Fig. 2(b)]: no $\beta$-crossed diffuse scattering is observed at $\phi$=90° while a cross-shaped feature with aperture $2\beta \approx 160°$ is measured at $\phi$=0°. In this case the average hillock apex angle measured by the AFM line scans along $\phi$=0° is 172°±2°, again not far from what deduced from the HR-RSMs.

In the HR-RSM recorded at $\phi$=90° from the 325°C sample [Fig.2(a)] we noticed an XDS halo with a characteristic intensity distribution along a specific direction defined by an angle $\gamma \approx 63°$. Moreover, this angle is not observed for HR-RSMs recorded with the primary X-ray beam along the perpendicular direction ($\phi$=0°). The above value of $\gamma$ coincides quite well with the 64.76° theoretical value of the angle between [113]- and [001]-directions in ZnTe, being thus clearly related to the asymmetric epilayer surface ridging observed in Fig.3(a) and Fig.4(a). It is worth noting that also here the AFM measurements and high-angle reciprocal space mappings are in excellent agreement.

## V. RELATIONSHIP BETWEEN SURFACE MORPHOLOGY AND LATTICE DEFECTS

A noteworthy result of our XDS analysis is the occurrence of SFs in the otherwise stress-free homoepitaxial ZnTe layers. Clearly, these SFs originate at the substrate-epilayer interface during the early stages of growth. The presence of residual trace oxides[10,21] and structural imperfections on the substrate surface may be indeed a source of SF nucleation. While no $\alpha$-crossed streaks are observed in the samples grown at 325°C, their intensity is more pronounced for $T_G \geq 375°C$ (Fig.2).

As the growth above 360°C leads to Te-rich surface conditions,[23] this seems to favor the SF nucleation; indeed, a sensitivity of SF nucleation towards VI-group atoms has been already reported in the case of ZnSe epilayers grown by MOVPE on GaAs.[24] Furthermore, the HR-RSMs reported in Fig. 2 evidence different intensities for the $\alpha$-crossed streaks recorded in the $\phi=0°$ and $\phi=90°$ sample setting, indicating different SF abundance along the two perpendicular <110> directions. This further implies that SFs lying along {111}Zn and {111}Te planes have different nucleation proclivity under the employed growth conditions.

The development of hillocks on the crystal surface is generally attributed to three-dimensional (3D) growth around spiral centers (screw dislocations) emerging at the surface, where the ad-atom migration on the surface is enhanced by an asymmetric potential barrier at the crystal step edges (Schwoebel effect).[25] However, the dislocation density in present ZnTe:P substrates is $<3\times10^4$ cm$^{-2}$, i.e. much lower than the hillock densities, estimated instead around $10^6$-$10^7$ cm$^{-2}$ (Ref. 23). This rules out the possibility that substrate dislocations propagating into the epilayers could reach the surface and act as source for hillock formation. The presence of SFs close to the epilayer-substrate interface is however, an effective source of 3D growth centers. Indeed, each SF is bound by two partial dislocations that, if emerging at the epilayer surface, may act as hillock nucleation centers, the expected material growth rate around them being higher than around screw dislocations under high supersaturation conditions (such as those holding in MOVPE).[26] As a matter of fact, almost all pyramidal hillocks on the epilayer surfaces appear in close pairs aligned along a common <110> direction [Fig. 3(c)], as expected when pairs of opposite partial dislocations (dipoles) bounding SFs act as 3D growth centers. The nucleation of hillock pairs along a preferential <110> direction further agrees with the different nucleation proclivity of SFs lying along {111}Zn and {111}Te planes, as discussed above.

Despite the XDS analysis of HR-RSMs did not allow a direct estimate of the epilayer SF densities, the correlation between SFs and hillock pairs on the epilayer surface does; on this basis, we estimate a SF density in the 375 °C and 400 °C samples of the order of $(4-5)\times10^6$ cm$^{-2}$.

A brief note should be made here as regarding the absence of hillocks on the surface of the 350 °C sample [Fig. 3(b)], despite the XDS analysis in the previous Section confirmed the presence of SFs. However, we have shown elsewhere[23] that pyramidal hillocks develop under Te-rich surface conditions, the 3D growth around a SF partial dislocation being driven by Te ad-atoms experiencing a Schwoebel potential barrier at the step edges around the center. Instead, almost featureless surfaces are obtained under nearly equal (stoichiometric) Te and Zn adatom abundances,[31] such as those applying to the sample in Fig.3(b). The morphology of ZnTe homoepitaxial layers grown at or above 350°C is therefore the result of a complex interplay between SF nucleation, ad-atom stoichiometry and Te surface diffusion mechanisms during MOVPE growth.

## V. CONCLUSIONS

The structural and morphological properties of MOVPE-grown homoepitaxial (001)ZnTe layers were investigated by X-ray diffuse scattering as function of the epilayer growth temperature. In particular, two diffuse scattering features were observed in the sample high-angle HR-RSMs that were related to (i) planar defects (SFs) within the ZnTe epilayers and (ii) morphological structure (pyramidal hillocks) at the epilayer surface. The characteristics of both XDS features are a cross-shaped diffuse scattering intensity around the (004) relp of ZnTe according to the following:

(i) Cross-shaped streaks defined by an aperture angle of $2\alpha \approx 70°$ were measured for the samples grown at $T_G$ >350°C. These $\alpha$–crossed streaks were not observed for the ZnTe:P substrate, demonstrating that such scattering features are strictly related to the epilayer. The orientation of the streaks in the RSMs corresponds exactly to the <111> crystallographic directions suggesting that SFs may be the origin of this diffuse scattering. Moreover, the The $\alpha$-crossed streak intensity increases with the epilayer growth temperature, indicating that the SF nucleation increases at high growth temperatures. Finally, the different intensity of the $\alpha$-crossed streaks recorded along the two <110> directions suggests a different proclivity of SFs to nucleate along {111}Zn and {111}Te planes.

(ii) Cross-shaped streaks defined by an aperture angle $2\beta \approx 160°$-$170°$ were instead measured for the samples grown above 360°C. The excellent agreement between the angle measured in the HR-RSMs and by AFM line scans demonstrates that the observation of the $\beta$-crossed streaks is the effect of pyramidal hillocks occurring on the epilayers under Te-rich surface conditions, their densities reaching up to $10^6$-$10^7$ cm$^{-2}$.

The mechanism of hillock formation has been briefly discussed and further ascribed to Te adatoms experiencing a Schwoebel potential barrier at the step edges around a 3D growth center. Pairs of partial dislocations (dipoles) bounding the SFs likely act as hillock nucleation centers in present samples, as demonstrated by the observation that hillocks appear on the epilayer surfaces in close pairs aligned along a common <110> direction.

Finally, the XDS analysis of the HR-RSMs recorded for the sample grown at $T_G$ =325°C (i.e. under kinetics limited conditions) evidenced a diffuse scattering halo with a characteristic intensity distribution defined by an angle $\gamma \approx 63°$, further attributed to an asymmetric ridging of the epilayer surface. The ridges are elongated along one of the <110> in-plane directions, their sidewalls being formed by {311} planes, as also confirmed by AFM observations.

**ACKNOWLEDGMENTS**

The authors would like to thank C. Minarini for her kind help during the X-ray diffraction measurements at CR ENEA of Portici (Italy) and M. Pentimalli for useful discussions. Drs. K. Sato and T. Asahi of Nikko Materials Co., Ltd. (Japan) are also acknowledged for supplying the ZnTe:P wafers used in this work. Special thank goes also to A. Pinna for his technical assistance during MOVPE growth experiments.

**FIGURE CAPTIONS**

**Figure 1:**

Reciprocal space maps recorded around the (004) relp from (a) the ZnTe epilayer sample grown at $T_G$ =350°C and (a) a ZnTe:P substrate. The ZnTe epilayer exhibits a diffuse scattering shape that is different from the circularly-shaped diffuse halo of the ZnTe:P substrate.

**Figure 2:**

(004) HR-RSMs of ZnTe epilayers grown at (a) 325 °C, (b) 375 °C and (c) 400 °C, for two different azimuth angular positions ($\phi$=0° and $\phi$=90°). Different features are observed: particularly, the extension of the X-ray diffuse scattering streaks along certain directions of the reciprocal space, corresponding to specific crystallographic orientations or surface morphology features.

**Figure 3:**

FEG-SEM micrographs of the surface morphology of the epilayers grown at (a) 325 °C, (b) 350 °C and (c) 400 °C. The surface morphology of the sample grown at 325°C shows ridges elongated along one of the <110>-directions, while the sample grown at 350°C shows almost no features. The sample grown at higher temperature exhibit pyramidal hillocks (density ~$10^6$-$10^7$ cm$^{-2}$) with their main rhombus-base diagonals aligned along a <110> direction.

**Figure 4:**

(a) An AFM topographic image of the epilayer grown at $T_G$ =325 °C [same sample as in Fig.3(a)]. A dense surface ridging is observed along a <110> direction. (b) The height profile obtained by an AFM line scan along the solid line shown in (a).

**Figure 5:**

Surface morphology of the epilayer grown at $T_G$ = 375°C as observed by AFM (a). The main diagonal of the pyramids are aligned along the <110> and $<\bar{1}10>$ directions as indicated by the arrows. These directions correspond to the sample setting $\phi$ = 0° and $\phi$ = 90° in the RSM measurements. AFM line profiles in (b) were recorded by the AFM tip moving along the directions $\phi$ = 0° and $\phi$ = 90° and yield the inclination angle of the sidewalls of the pyramids.

**Figure 6:**

Scheme of the reciprocal space mapping in vicinity of the ZnTe(004) reciprocal lattice point in the real (a, c) and reciprocal space (b, d) for the experimental observations of the structural defects (a, b) and morphologial features (pyramids, ripening) on the ZnTe epilayer surface (c, d). The crystallographic directions and the angle between the experimentally observed diffuse x-ray streaks with respect to the sample surface (in-plane direction <110>) and surface normal (<001> direction) are indicated.

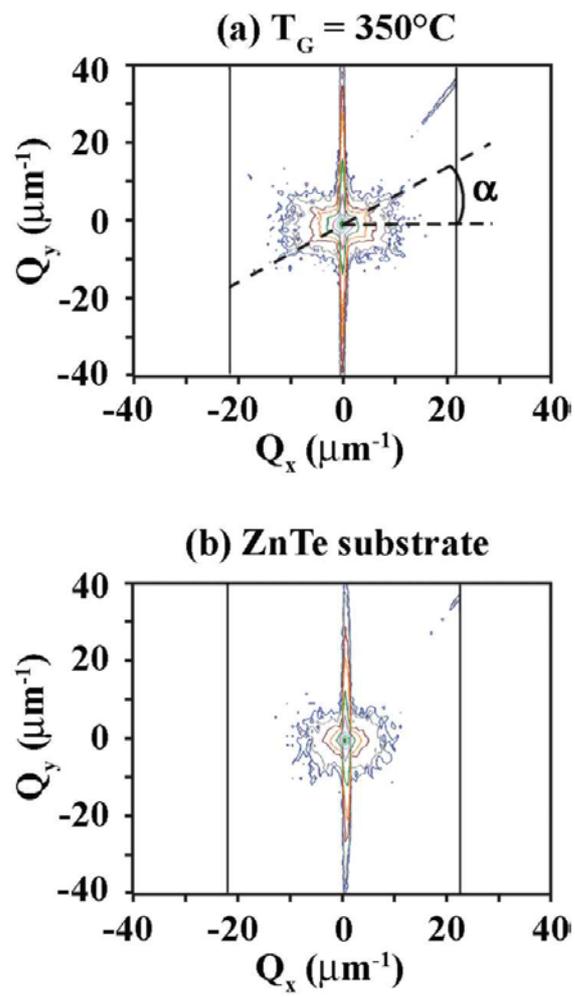

**Figure 1**

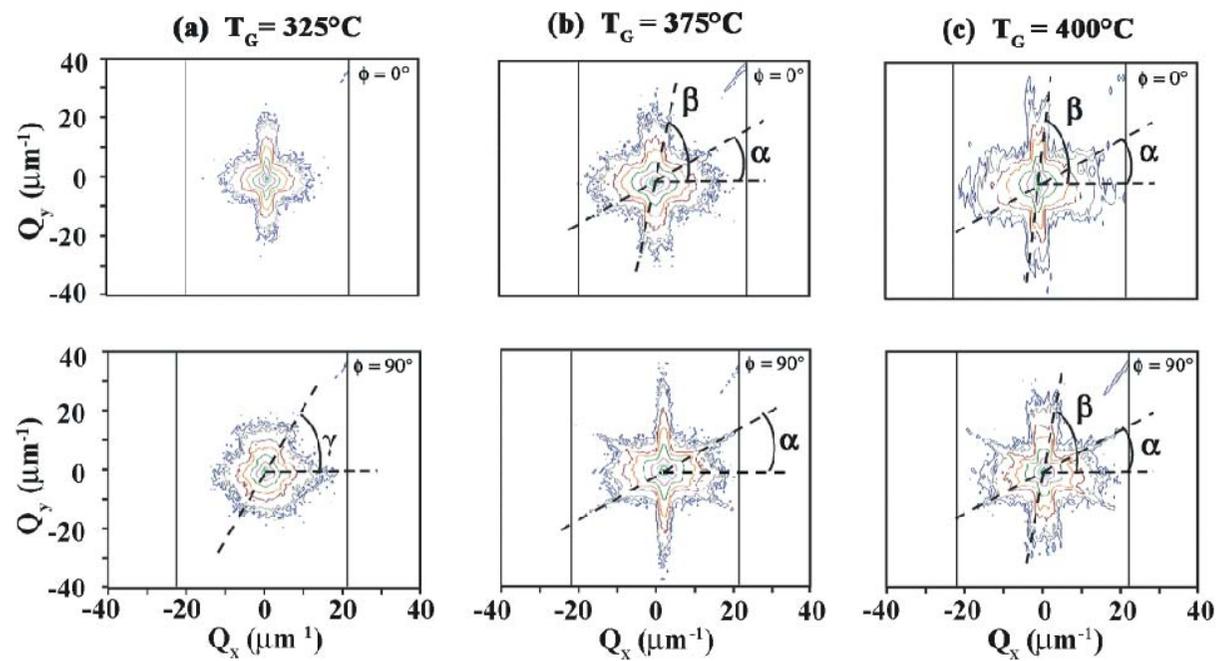

**Figure 2**

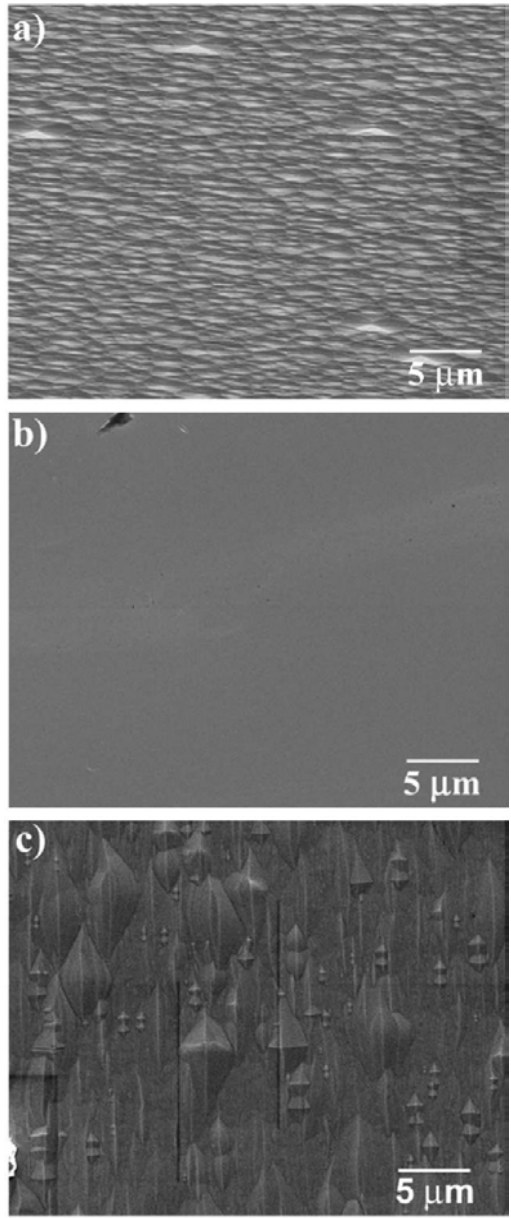

**Figure 3**

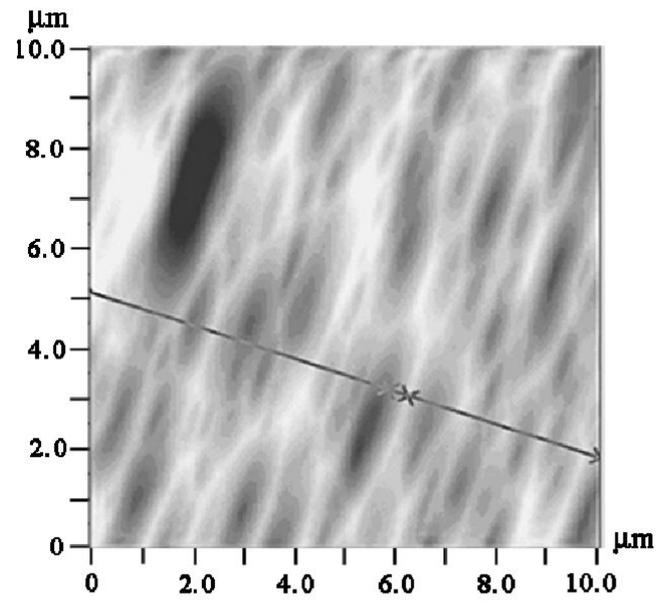

(a)

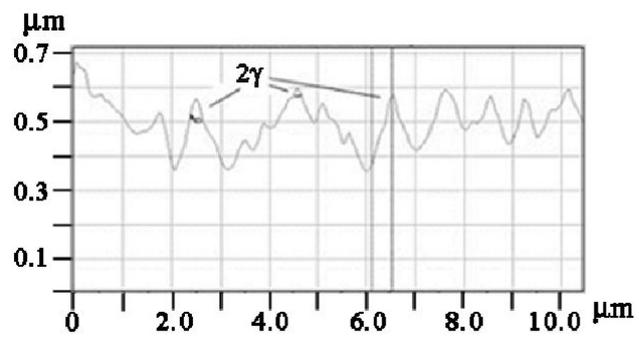

(b)

**Figure 4**

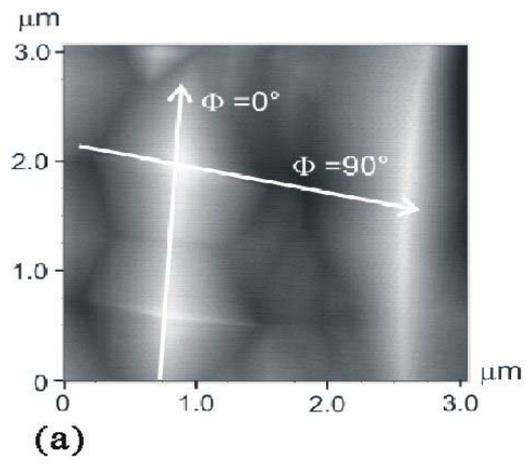

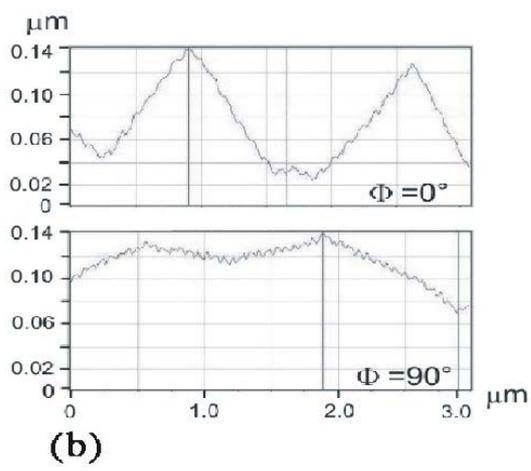

**Figure 5**

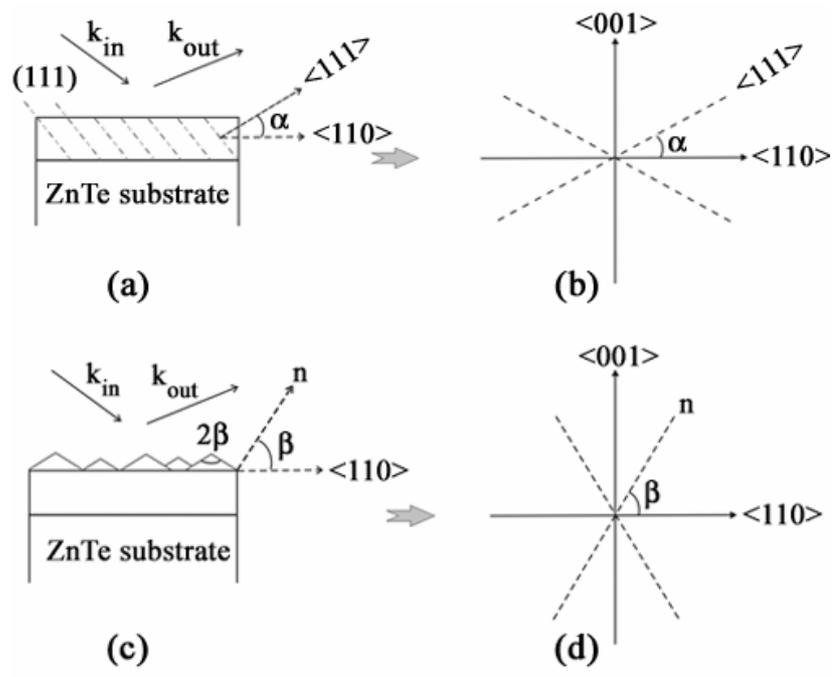

**Figure 6**